# Analysis of Fractional Vegetation Coverage and its Dynamic Change in the Yalong River Basin based on Dimidiate Pixel Model

Yue Qin*, Yuwei Lyu

*State Key Laboratory of Water Resources and Hydropower Engineering Science, Wuhan University, Wuhan 430072, China*

*Corresponding author (E-mail: AmberQ@whu.edu.cn)

**Abstract**
Fractional vegetation coverage (*FVC*) and its spatio-temporal variations are critical indicators of regional ecological changes, which are of great significance to study the laws of surface variation and analyze regional ecosystem. Under the development of RS and GIS technology, this analysis employs Landsat satellite images in 1994, 2008, 2013 and 2016 to estimate *FVC* in Yalong River Basin based on the Dimidiate Pixel Model. With consideration of the vegetation coverage condition and land surface law in the study area, the research further analyzes the Spatio-temporal variations as well as the influencing factors of FVC in terms of topography and land use types respectively. The results show that since 1994, *FVC* in Yalong River Basin has experienced a downward trend yet displaying an uptick from 2013. Moreover, different land use types indicate the versatility of land covers in Yalong River Basin, with grassland and forest performing probably the most important factors that can induce changes to the stability of *FVC* in whole basin. Overall, the research reflects the impact of human activities on vegetation in Yalong River Basin, and provides available data and theoretical basis for ecological assessment, ecological restoration and environmental protection.

## 1. INTRODUCTION

The Yalong River Basin, located in the southwestern Sichuan Province, is a habitat of native plants and is the largest tributary of Jinsha River that has plentiful animal and plant communities. With an abundance of hydropower, mineral and forestry resources, however, Yalong River Basin has been under immense threats caused by human irrational exploitation and intervention over decades, including habitat degradation, ecological recession, water loss, soil erosion, geologic hazards, etc.

Fractional vegetation coverage (*FVC*) is a critical climatic and ecological parameter, also a composite indicator to measure plant cover. Hence, it is important to conduct research on FVC for the analysis and evaluation on regional eco-environmental changes. Furthermore, *FVC* and its change would exercise a crucial influence on hydrology, ecology and global changes.

Those years, when China is placing emphasis on ecological civilization, have witnessed flourishing researches on the ecological succession, evaluation and restoration of Yalong River Basin. For instance, Tian et al. employed the method of trend line analysis with correlation analysis to evaluate the spatial and temporal variations of vegetation coverage during a 24-year span (1982 to 2006) in Yalong River Basin[1]. Wang et al. analyzed the overall changes and spatial characteristics of the ecological environment by the means of remote sensing and GIS on Yalong River Basin[2].

There are two types of data that can be utilized to estimate vegetation coverage: observed data and remote sensing image data. With the quality limited by certain objective conditions, observed data is often applied to evaluate vegetation coverage on a small scale. While the dimidiate pixel model using remote sensing image data is now a common method of measuring regional plant cover, Guo et al. applied MODIS data to estimate the vegetation coverage in arid zones [3]; Long et al. carried out a research on karst rock-desertification using NDVI and dimidiate pixel model [4]; Jiang et al. calculated plant cover in Anding District, Dingxi City of year 1999 and 2007 respectively, and demonstrated its change between the two years based on the same model [5].

Due to the absence of accessible research results on the vegetation change in Yalong River Basin from 2010 onward, we individually evaluate FVC in the area of 1994, 2008, 2013 and 2016, then referring to existing relevant information, we comprehensively analyze the spatial and temporal changing patterns of plant cover in terms of interannual variation, topography, and land use types, with focus on the underlying reasons. The general objective of our study is to reflect the impact of human activities on vegetation in Yalong River Basin for further purpose of providing available data and theoretical basis for ecological assessment, ecological restoration and environmental protection.

## 2. STUDY AREA AND DATA

*2.1 Study Area*

Yalong River, the largest first-order tributary of Jinsha River located in upper Yangtze River, with a main stem at the length of 1571 km, originates from the Bayan Har Mountains in Qinghai province, running southeastward through western Sichuan Province and confluencing with Jinsha River in Panzhihua City. Yalong River Basin (26°32′~33°58′N, 96°52′~102°45′E) is a long thin catchment, high in the north, west and east, sloping and expanding southward.

Yalong River Basin covers approximately 136 thousand km², including pasture of 43.28 thousand km², forest around 19.29 thousand km² and cultivated lands of 0.193 thousand km². The region is inhabited by over two million people. To be specific, the upper and middle basin is less populous and more spacious, as one of the animal husbandry bases in Sichuan, while the lower part is densely populated with advanced industry and agriculture.

In Yalong River Basin, the average annual precipitation, evaporation and temperatures range from 500 mm to 2470 mm, 1166 mm to 2500 mm and -4.9℃ to 19.7℃ selectively. The three indicators above increase from north to south, while the average annual humidity varies between 53% and 73%, not showing much difference along the basin.

Yalong River main stem carries hydropower of around 33.4 million KW with 21 cascade hydropower stations are being planned. There are lots of dams completed or under construction, inclusive of Jinping 1 Dam, Jinping 2 Dam, Guandi Dam, and Ertan Dam, which was the largest hydropower station in the late 20th century in China, located in the lower course of Yalong river.

*2.2 Data and Preprocessing*

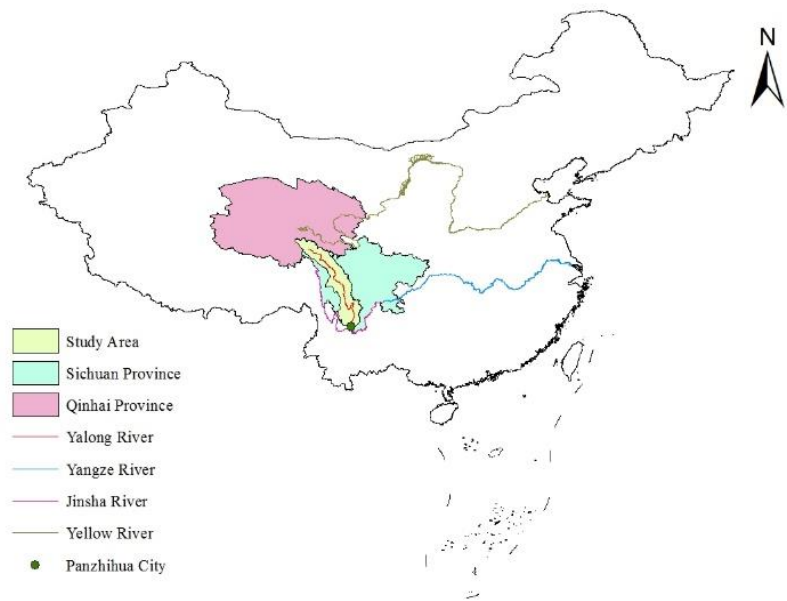

**Figure. 1** Location of Yalong River Basin in China

The main data source in our study are multispectral remote sensing images from satellite Landsat 5 and Landsat 8. We applied cloudless images of year 1994, 2008, 2013 and 2016 in summer. After preprocessing them with ENVI by radiometric correction, geometric correction and etc., we further mosaicked and clipped the data as needed, then used atmospheric correction to decrease effects of the aerosol.

### 3. METHODOLOGY

*3.1 Dimidiate Pixel Model*

Being one typical mixed-pixel decomposition model, the dimidiate pixel model is a preferred model for estimating *FVC*. Accordingly, we adopt this model in the study of *FVC* in Yalong River Basin. The dimidiate pixel model is an easy yet practical remote sensing model, which assumes that a pixel consists of two parts: pure vegetation and non-vegetation. Also, the spectral statistics observed by remote sensor are a weighted linear combination of the two components, each of whose weight equals the proportion of their area respectively (e.g., *FVC* as the weight of vegetation) [6].

Based on the above theory, the information *S* observed by remote sensor will be given by

$$S = S_{veg} + S_{soil} \qquad (1)$$

where $S_{veg}$ stands for the information provided by green vegetation and $S_{soil}$ is the information provided by non-vegetation.

Now suppose that the proportion of the area where a pixel is covered with vegetation is *FVC* (i.e., the vegetation coverage of this pixel), then the proportion of the area without vegetation is 1- *FVC*. If what a purely vegetation-covered pixel provides is the remotely sensed information $S_{veg}$, the information contributed by vegetated part of the mixed pixel ($S_v$) can be calculated therefore according to

$$S_v = FVC \cdot S_{veg} \tag{2}$$

Similarly, if a pure pixel covered thoroughly with bare soil provides the remotely sensed information $S_{soil}$, the information contributed by soil in the mixed pixel ($S_v$) can be described as

$$S_s = (1 - FVC) \cdot S_{soil} \tag{3}$$

Rewrite Equation (1) with parameters in Equation (2) and Equation (3). Thus *S* is expressed by

$$S = FVC \cdot S_{veg} + (1 - FVC) \cdot S_{soil} \tag{4}$$

Transform the equation above, we can consequently describe *FVC* as

$$FVC = (S - S_{soil}) / (S_{veg} - S_{soil}) \tag{5}$$

where $S_{veg}$ and $S_{soil}$ are two parameters of the dimidiate pixel model.

Hence, with the two parameters we can apply remote sensing information to estimate *FVC* based on Equation (5).

The model in question exhibits the relation between remote sensing information and *FVC*, with two parameters ($S_{veg}$ and $S_{soil}$) of practical significance. In this sense, the model can diminish the impact of atmosphere, background of soil, vegetation type and etc., thus minimizing their effects on the remote sensing information [7].

*3.2 Normalized Difference Vegetation Index (NDVI)*

Vegetation Indices are unique signals extracted from the optical properties of canopies, also a simple and effective remote sensing parameter to measure vegetation cover on the ground surface and its growth condition [8].

Normalized Difference Vegetation Index (*NDVI*) is being widely applied in remote sensing of vegetation. As the optimal indicator of the growing conditions for vegetation and its coverage, *NDVI* is linearly correlated with the vegetation density that can be used to quantify biomass and vegetation detection. Experiments have been carried out to demonstrate that the application of *NDVI* is more suitable for detection of vegetation in the middle of its growth stage or with a medium coverage.

The data in our analysis was collected in September, we thus employed *NDVI* to quantify *FVC* in the Yalong River Basin. *NDVI* is calculated by the following equation:

$$NDVI = (NIR - R) / (NIR + R) \tag{6}$$

where *NIR* and *R* represent the spectral reflectance measurements acquired in the near-infrared and red regions selectively.

*NDVI* falls in a range from −1 to 1, with negative values suggesting a larger reflectance of *R* yet higher absorption of *NIR* by a certain object, and vice versa. While the former ones represent objects like clouds, water, snow, etc., the positive values indicate areas with *FVC* which increases as *NDVI*. The approximate equality of *R* and *NIR* will result in a *NDVI* value of 0 that corresponds to areas of rock or bare soil.

Some factors, such as the solar elevation angle, topography and shadow, which have impacts on satellite images, can produce abnormal *NDVI* values (greater than 1 or less than 0) in few cases. Therefore, to remove those abnormal results ahead of the estimation, we assume the negative values to be zero serving as background value, and give those higher than the upper limit exactly a value of one likewise. Thus, we transform the band values into unsigned ints.

*3.3 Estimation of FVC Using NDVI*

Based on *NDVI* we employ the pixel decomposition model to estimate *FVC* as a scaled *NDVI*

$$FVC = (NDVI - NDVI_{soil}) / (NDVI_{veg} - NDVI_{soil}) \tag{7}$$

where $NDVI_{soil}$ is the *NDVI* of bare soil or areas under no vegetation, and $NDVI_{veg}$ is the *NDVI* for pure vegetation.

Assuming $NDVI_{soil}$ and $NDVI_{veg}$ respectively as the minimum and maximum *NDVI* from our study area within the confidence interval at a given confidence level respectively, the algorithm in Equation (8) thus can be derived:

$$FVC = (NDVI - NDVI_{min}) / (NDVI_{max} - NDVI_{min}) \tag{8}$$

## 4. APPLICATION AND RESULTS

### 4.1 Classification of Land Cover Data

The land cover types of the study area fall into five categories: forest, farmland, water, residential area and other land. We use the processed images, band combination and classified values of ISODATA to generate multivariate data. Then we utilize the samples extracted by the classification workflow to perform supervised classification on the data. The exact supervised classification method in our analysis is Neural Net Classification.

### 4.2 Acquisition of NDVI

Theoretically, $NDVI_{soil}$ is a constant (zero) that seldom varies. Whereas, when it comes to different soil types, the figure will fluctuate from -0.1 to 0.2[9]. The same is true for $NDVI_{veg}$ value as it can change with various land cover types and vegetation phenological period. Therefore, higher accuracy can be achieved when selection of $NDVI_{soil}$ and $NDVI_{veg}$ is on the basis of land cover.

With the land use images serving as a reference, we assume the minimum and maximum cumulative frequency within confidence intervals (5%~95%) to be $NDVI_{soil}$ and $NDVI_{veg}$ respectively.

According to Tian's research[1], the decade from 1990 to 2000 sees a low $NDVI$ value along with slight fluctuations in Yalong River Basin. $NDVI$ hits the bottom during 1991 to 1993, a period in line with the time when Ertan Dam was under construction (probably something to do with deforestation). While from year 2001 onward, $NDVI$ value has displayed an upward trend. All the conclusions above are consistent with our results.

**Table. 2** *NDVI* of Different Years and Land Cover Types

| Year<br>Type | 1994 | | 2008 | | 2013 | | 2016 | |
|---|---|---|---|---|---|---|---|---|
| | $NDVI_{soil}$ | $NDVI_{veg}$ | $NDVI_{soil}$ | $NDVI_{veg}$ | $NDVI_{soil}$ | $NDVI_{veg}$ | $NDVI_{soil}$ | $NDVI_{veg}$ |
| Forest | 0.4590 | 0.8040 | 0.6392 | 0.8588 | 0.6373 | 0.8196 | 0.6403 | 0.8972 |
| Farmland | 0.3680 | 0.8230 | 0.0039 | 0.6157 | 0.4714 | 0.6435 | 0.3581 | 0.6792 |
| Water | 0.0000 | 0.0000 | 0.0000 | 0.0000 | 0.0000 | 0.0000 | 0.0000 | 0.0000 |
| Residential Area | 0.6700 | 0.7730 | 0.1137 | 0.6627 | 0.1550 | 0.5015 | 0.0037 | 0.2523 |
| Other Land | 0.6430 | 0.7640 | 0.0667 | 0.3804 | 0.3712 | 0.6432 | 0.4628 | 0.7588 |

Note: theoretically, water has a NDVI value of 0.

### 4.3 Estimation of FVC and Display of Images

We derive *FVC* of four years (1994, 2008, 2013, 2016) from Equation (2), and display the results in an intuitive way by density slicing.

Based on spectral characteristics of the pixels, we categorize *FVC* into 5 classes: extremely low *FVC* (0%~20%), low *FVC* (20%~40%), medium *FVC* (40% ~ 60%), high *FVC* (60% ~ 80%) and extremely high *FVC* (80%~100%). To provide the basis for quantitative analysis on the spatial distribution and changing pattern of *FVC* in Yalong River Basin, we finally generate four images (Fig.2), Table. 2 and Table. 3.

### 4.4 Interannual Variation of FVC

#### 4.4.1 Comparison of Spatial Distribution

As demonstrated in Fig. 2, from 1994 to 2016, *FVC* of Yalong River Basin is overall in a downturn that is remarkable in the upper basin while slightly in the lower basin. Again, this trend is in line with Tian's research[1]. For the possible causes, we deem that the decrease of *FVC* in the upper basin is relevant to excessive exploitation of forest resources, in the lower basin allied to the construction of hydropower project and climatic variation.

#### 4.4.2 Interannual Variation

**Table. 2** Rate of *FVC* in Yalong River Basin

| Year | Percentage of Area Under Vegetation | | | | |
|---|---|---|---|---|---|
| | 0%~20% | 20%~40% | 40%~60% | 60%~80% | 80%~100% |
| 1994 | 4.92 | 12.46 | 23.34 | 19.12 | 40.16 |
| 2008 | 7.49 | 5.67 | 21.04 | 27.20 | 38.60 |
| 2013 | 10.08 | 14.22 | 18.84 | 26.53 | 30.34 |
| 2016 | 10.84 | 19.68 | 19.55 | 13.91 | 36.01 |

In Fig. 2 and Table. 2, areas of extremely low *FVC* (0%~20%) and low *FVC* (20%~40%) displayed increase during the 22-year span in varying degrees; areas of medium *FVC* (40%~60%) flattened out; proportion of the area under high *FVC* (60%~80%) rose first and declined later, with drastic drop from 2013 to 2016 in particular. Conversely, areas covered by dense vegetation (with *FVC* of 80%~100%) reduced gradually before 2013, whereas increasing in 2016.The rate of change in *FVC* also varies with different levels. From 1994 to 2008, the highest rate of change (53.88%) could be found in areas under high *FVC*, with the least one (-10.4%) in cases of extremely high plant cover. In the next stage, from 2008 to 2013, the rate of change ranked first in regions of low *FVC* at the level of 14.2%, whereas it ranked last (-11.2%) in highly vegetated areas. In the latest period (2013 to 2016), the maximum and minimum rate of change were in proportion of areas under high and medium *FVC*, -31.5% and 17.9% respectively.

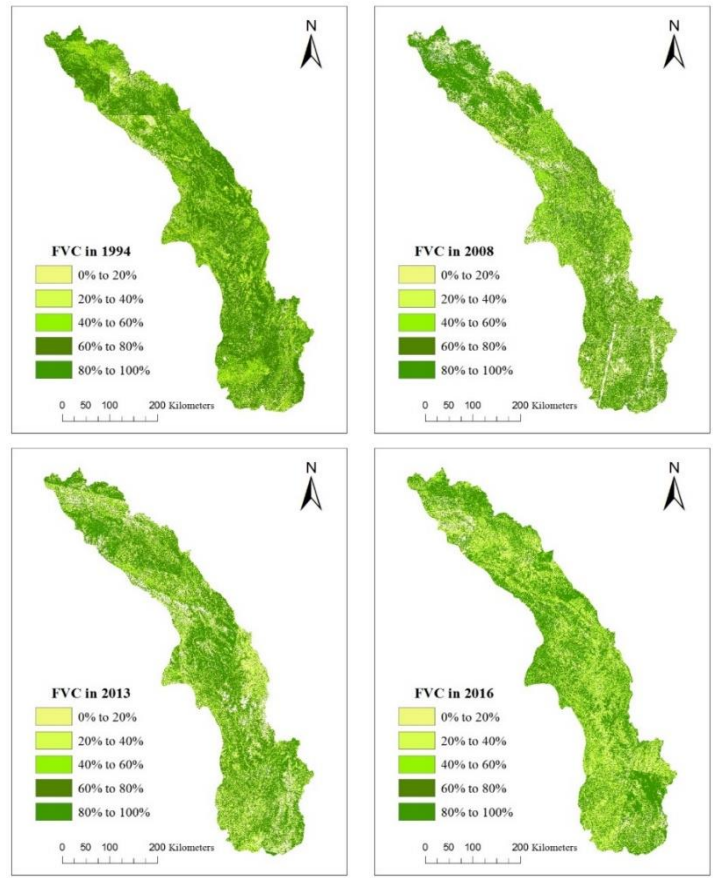

**Figure. 2** Rate of *FVC* in Yalong River Basin

*4.4.3 Overall Analysis*

Table. 3 exhibits a general trend that, during the period in our research, average *FVC* ($FVC_{mean}$) in Yalong River Basin declined gradually yet displayed an uptick in the last period, with the standard deviation (Std Dev) of *FVC* at a relatively stable level of around 0.30.

**Table. 3** Interannual Variation of $FVC_{mean}$ in Yalong River Basin

| Year | 1994 | 2008 | 2013 | 2016 |
|---|---|---|---|---|
| $FVC_{mean}$ | 0.64 | 0.56 | 0.54 | 0.58 |
| Std Dev | 0.29 | 0.34 | 0.33 | 0.34 |

Specifically, $FVC_{mean}$ is comparatively high at 0.64 in 1994: this fact may be explained by the relatively small-scaled impacts of human interventions on the vegetation in early stages. Nonetheless, the value then underwent an approximately linear decrease and bottomed out in 2013 at 0.54. Chances are that urbanization and population boom have led to the surge in demand for food, fuel, etc., which promote abusement of natural resources in Yalong River Basin, and lead to the decrease in plant cover to a certain extent. Dramatically, $FVC_{mean}$ rise slightly to 0.58 from the year 2013 onward, probably due to the fact that growing emphasis on environmental issues results in the implement of plant conservation and recovery. Moreover, the estimation that the average *FVC* remains about 0.55 in 2008, 2013 and 2016 reveals that the plant cover in Yalong River Basin remains steady during recent years.

Meanwhile, the standard deviation (Std Dev) of *FVC* in four years demonstrates no significant change: the year 1994 witnessed a minimum Std Dev at 0.29, indicating small variation with regard to *FVC* rates in that phase. Afterward, the value which flattened out (approximately 0.34) in the next three periods reveals the slightly widened gap between the plant cover in different parts of Yalong River Basin compared with that in late 20th century yet already stayed at a stable stage.

### 4.5 Impact of Topography on FVC

We divide the elevation data of Yalong River Basin with ArcGIS into five classes: 969 m ~2500 m , 2500 m ~3500 m, 3500 m ~4500 m, 4500 m ~5500 m, 5500 m ~5876 m (Fig. 3 ). Then masks are created accordingly in ENVI to compute $FVC_{mean}$ under each elevation range. Fig. 4 depicts the average *FVC* of Yalong River Basin in 1994, 2008, 2013 and 2016 categorized by elevation.

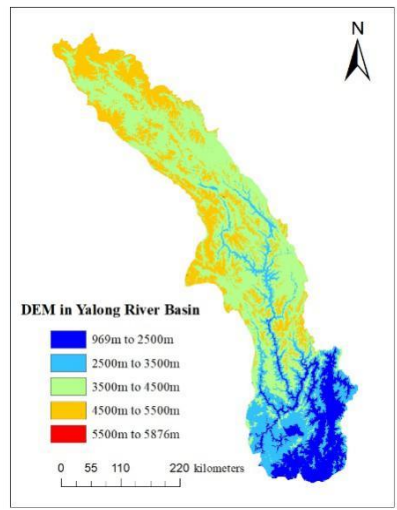

**Figure. 3** DEM (Digital Elevation Model) in Yalong River

Fig. 4 indicates a high *FVC* in Yalong River Basin: areas under medium and higher plant cover at each elevation range all accounted for over 50%. Major part of vegetation is in areas at above 5500 m and high *FVC* takes up the largest proportion, greater than that at other elevation significantly. The international standard defines elevation of over 5500 m as extreme altitude where alpine sparse vegetation is common. From other study[2] we acquire the knowledge that the upper Yalong River Basin on a plateau experiences a temperate semi-arid climate with little precipitation. Therefore, we deduce that *FVC* may well be affected by the climate conditions in regions at high altitude, such as global warming, melting icebergs, etc., which brought about the noticeable increase in *FVC*.

In terms of the interannual variation, areas at low elevation (lower than 2500 meters) took up a growing percentage under high and extremely high *FVC* from 1994 to 2008, given the fact that Sichuan carried out plans to restore arable land (805.3 thousand ha) to nature during the period of 1999 to 2004, with some regions along Yalong River (e.g., Ganzi County, Liangshan Prefecture and Panzhihua City) serving as the core area. On top of that, Sichuan became the first in China to ban deforestation in 1998. However, due to the interruption of implementation of the policy with reinforcement of economic development after year 2008, areas of high and extremely high plant cover reduced, yet less vegetated areas took up larger proportion accordingly. In Wang.'s research[2], regions of Yalong River Basin that suffer from ecological degradation are primarily situated in Ganzi County, Yanyuan County, and places from the lower basin which feature in abundant mineral resources, advanced economic performance, dense population and much human intervention. Hence, it is inferred that human intervention displays a significant role in plant cover of areas at low altitude.

As a whole, given some relevant researches, factors that have impact on *FVC* in Yalong River Basin with regard to topography are human intervention and climate conditions.

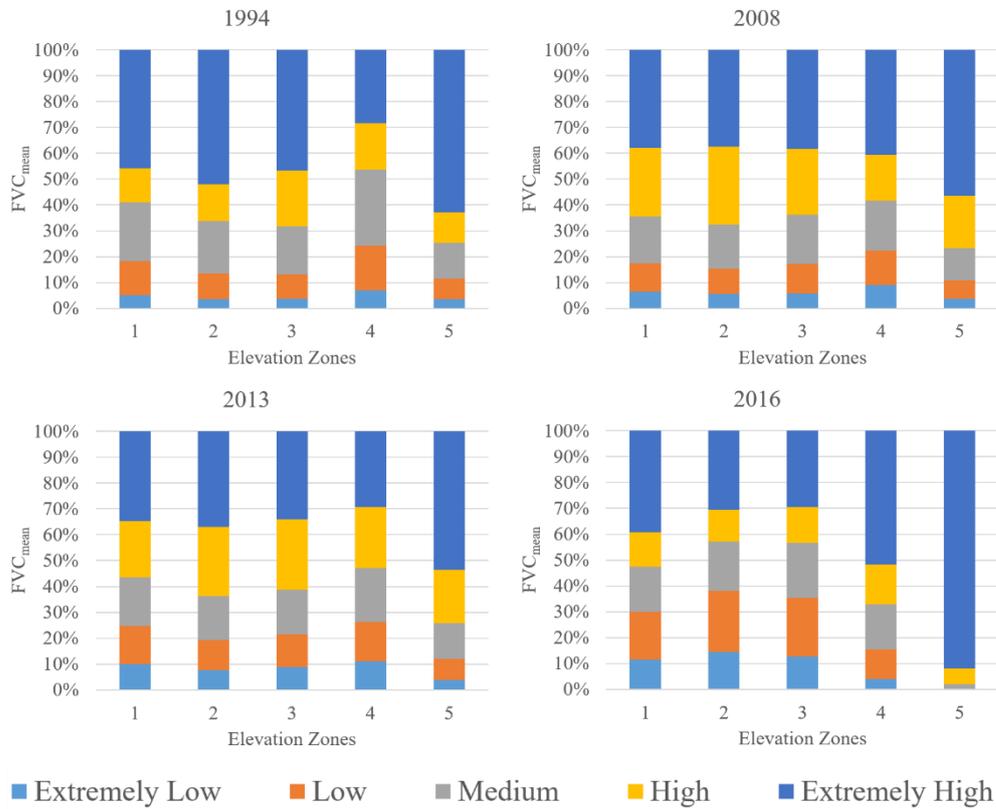

**Figure. 4** $FVC_{mean}$ Classified by Elevation in Yalong River **Basin**

### 4.6 Impact of Land Use Types on FVC

From the Institute of Geographic Sciences and Natural Resources Research, CAS, we obtain available land-use data of year 1994, 2008 and 2013, then extract information on Yalong River Basin and reclassify it into six categories: farmland; forest; grassland; water; urban, rural, industrial and residential area; unused land. All above serve as a data supplementation for our analysis on *FVC*.

**Table. 4** Proportion of Area Classified by Land Use Types in Yalong River Basin (%)

| | Land Use Type | Year | | |
|---|---|---|---|---|
| | | 1994 | 2008 | 2013 |
| 1 | Farmland | 5.43 | 5.42 | 5.31 |
| 2 | Forest | 35.26 | 35.14 | 36.26 |
| 3 | Grassland | 51.53 | 51.57 | 51.17 |
| 4 | Water | 0.49 | 0.58 | 0.63 |
| 5 | Urban, Rural, Industrial and Residential Area | 0.12 | 0.12 | 0.20 |
| 6 | Unused Land | 7.17 | 7.17 | 6.43 |

The statistics in Table. 4 demonstrate some changes in the percentage of area in different land uses. During the first period (1994~2008), the proportion of area from most categories remained relatively stable, with the greatest change occurred in forest (-0.12%). At the next stage (2008~2013), however, the percentage in the case of forest experienced a noticeable increase (1.12%); meanwhile, the figure of farmland reduced by 0.11% and that of grassland dropped by 0.4%, still occupying the largest proportion (over half of our study area).

**Table. 5** $FVC_{mean}$ Classified by Land Use Type in Yalong River Basin

| Land Use Type | $FVC_{mean}$ | | |
|---|---|---|---|
| | 1994 | 2008 | 2013 |

| | | | | |
|---|---|---|---|---|
| 1 | Farmland | 0.56 | 0.44 | 0.13 |
| 2 | Forest | 0.74 | 0.61 | 0.49 |
| 3 | Grassland | 0.61 | 0.57 | 0.45 |
| 4 | Water | 0.47 | 0.22 | 0.02 |
| 5 | Urban, Rural, Industrial and Residential Area | 0.46 | 0.35 | 0.01 |
| 6 | Unused Land | 0.47 | 0.34 | 0.12 |

Table. 4 and Table. 5 suggest a certain degree of the relation between land use types and *FVC*. As the temperate semi-arid climate prevails in upper Yalong River Basin with little precipitation, the dominant vegetation is grassland. Along the middle and lower reaches of Yalong River, some regions are rich in mineral resources, high in the level of economic development and are densely inhibited. When it comes to south-eastern part of the lower basin, monsoons bring sufficient water and heat to humid regions which are mainly covered by indigenous subtropical evergreen broad-leaved forests and temperate deciduous broad-leaved forests[2].

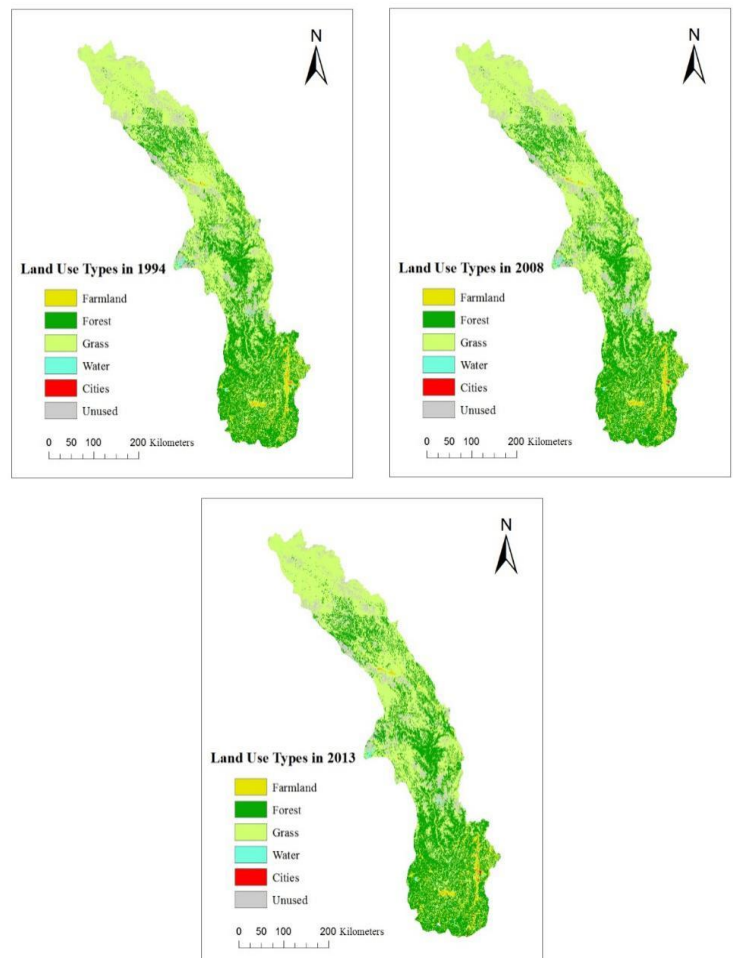

**Figure. 5** Land Use Types in Yalong River Basin

Over the periods, *FVC* value of forest was higher than that of grassland, while farmland had a smaller *FVC* than grassland did; plant cover in urban, rural, industrial and residential areacomparatively stayed at a low level. All results above are in accordance with the common sense.

*4.7 Possible Causes*

*4.7.1 Natural Factors*

91.5% of Yalong River Basin sits in Sichuan, a province that is prone to earthquakes. From 1999 to 2008 the Yajiang earthquake occurred with magnitude of 6.0; the period (2008~2013) saw three earthquakes at magnitude over 6.0, including the most devastating Wenchuan earthquake. Some researches[10] [11] have demonstrated the impact of earthquakes on forestry resources, thus reducing plant cover.

*4.7.2 Human Factors*

The Google Map illustrates a growing density of buildings and highways in urban, rural, industrial and residential area in 2013 compared to that in 2008, which might account for the drop of *FVC* in this case. Additionally, the exploitation of plants in Ganzi County, Liangshan Prefecture and Panzhihua City along Yalong River would result in decreasing *FVC* as well.

## 5. DISCUSSION

*5.1 Error Analysis*

Given the low resolution in remote sensing images, it is difficult to identify bare soil, farmland and residential areas when selecting samples with ROI tools. Some subjective factors also led to sampling errors. Furthermore, due to inaccurate classification, the masks that we later created could not present the actual range of data from different land covers.

Another contribution to errors is the cloud cover. Despite the efforts we made to select cloudless images, it was inevitable to work with cloud background given that we had to mosaic 12 satellite images to generate a full picture of our study area. For instance, cloud covered 61.58% of the remote sensing image on September 18, 2013, and the cloud amount in an image on July 14, 2008 was 98.29%.

Additionally, factors including surface albedo, vegetation spectrum that are affected by zenith angle, observation angle soil background would interfere with *FVC* values when calculated by *NDVI*[12].

*5.2 Method Comparison*

After pre-processing the Landsat images, we reached the stage of estimating *FVC*. With neural network classification (supervised), we categorized the land cover in Yalong River Basin into five classes. Thereafter, we employed the dimidiate pixel model to evaluate overall *FVC* of our study area.

Here we focus our attention on the advantages and disadvantages of neural network classification and dimidiate pixel model.

*5.2.1 Neural Network Classification*

Based on a nonparametric classifier, neural network classification features highly nonlinear classification, distributed processing, high accuracy, etc. It can eliminate the ambiguity and uncertainty of conventional classification methods in the domain of remote sensing at a certain degree.

Nonetheless, human errors will contribute to inaccuracy when selecting representative samples of land cover; it is also very time-consuming [13] in order to better perform classification that comes with more iterations. One research[14] suggests that, among four commonly applied techniques (minimum distance, maximum likelihood, neural network and support vector machine), neural network classification is relatively more accurate but consuming the longest time.

*5.2.2 Dimidiate Pixel Model*

The dimidiate pixel model assumes that two components (pure vegetation and non-vegetation) make a pixel, while remote sensing information is exactly a linear combination of the spectral signal of vegetation and soil with the proportion of their area serving as respective weight. The model utilizes NDVI as the sole parameter to calculate FVC over the entire basin without additional estimation of other vegetation indices like *LAI* (leaf area index). Hence, with consideration of the fact that it is easy to understand the theory and bring it to practice, dimidiate pixel model is rather a practical model for estimating *FVC*.

Limitations (e.g., low resolution of satellite images, cloud amount and the catchment factors of our study area) still make it hard to find out pixels entirely under plants or bare soil, though. In addition, *NDVI* will reach its saturation point when *FVC* is over 80% thereby will not change with the vegetation density any longer [18]. From this perspective, significant uncertainty lies in the evaluation of *FVC*.

## 6. COUCLUSION

*6.1 Comprehensive Analysis of FVC in Yalong River Basin*

Employing the combination of supervised classification and dimidiate pixel model, we find that since 1994, *FVC* in Yalong River Basin has experienced a downward trend yet displaying an uptick from 2013, where human intervention (i.e., deforestation, construction of hydropower projects, exploitation of mineral resources and implementation of relevant policy) is playing a key role. Moreover, different land use types present the versatility of land covers in Yalong River Basin, with grassland and forest performing probably the most important factors that can induce changes to the stability of *FVC* in the whole basin.

For the purpose of strengthening the long-term development and maximizing the economic benefits in Yalong River Basin, we suggest that in the process of natural resource planning, environmental protection, government should place greater emphasis on the integrity and stability of grassland and forest, with highlighting soil/ water conservation and vegetation recovery.

*6.2 Comprehensive Analysis of the Method*

Recently, given the characteristics including short period, high speed, large area coverage, effective radar seldom affected by weather, etc., the technology of remote sensing is considered as an advanced method with hyper spectrum and high spatial/ temporal resolution, and it has been critical in the dynamic monitoring of ecological changes.

On the other side, in practical application of estimating *FVC*, the dimidiate pixel model inevitably remains some disadvantages, e.g., limitations of satellite images, inaccurate NDVI values interfered by the surface albedo, vegetation spectrum and etc., still. These factors make the method in need of improvement.